\documentclass{emulateapj}
\usepackage{natbib}

\shorttitle{IC~10 with LGSAO}
\shortauthors{Vacca et al.}

\begin{document}

\title{Imaging of the Stellar Population
of IC~10 \\
with Laser Guide Star
Adaptive Optics \\
and the {\it Hubble Space Telescope}}

\author{William D. Vacca\altaffilmark{1},
Christopher D. Sheehy\altaffilmark{2} , and 
James R. Graham\altaffilmark{2,3} } 
\affil{Dept.\ of Astronomy, University
of California, Berkeley, CA 94720-3411} 

\journalinfo{ApJ in press; submitted August 17, 2006: accepted January 19, 2007}

\altaffiltext{1}{SOFIA-USRA, NASA Ames Research
Center, MS N211-3, Moffett Field, CA 94035; wvacca@sofia.usra.edu}
\altaffiltext{2}{Center for Adaptive Optics, 
University of California, Santa Cruz, CA 95064, U.S.A. }
\altaffiltext{3}{jrg@berkeley.edu}

\begin{abstract}

\noindent

We present adaptive optics (AO), near-infrared images of the central
starburst region of the Local Group dwarf irregular IC~10.  The Keck 2
telescope laser guide star facility was used to achieve near
diffraction limited performance at $H$ and $K'$ with Strehl ratios of
$18$\% and $32$\%, respectively.  The images are
centered on the putative Wolf-Rayet (W-R) object [MAC92]~24.  
Photometry from AO images can be subject to large systematic errors
($\simeq 1 $ mag.) due to uncertainties in the point spread functions
(PSFs) and the associated encircled energy curves caused by Strehl
variations.  However, IC~10 presents a rich star field, and therefore
we are able to use the Fourier power spectrum method of Sheehy,
McCrady and Graham (ApJ, 647, 1517) to reconstruct the photometric
curve of growth from the data themselves, and thereby reduce the
magnitude of the systematic errors in our photometry to $\leq 0.04$
mag.
We combine our ground-based images with an F814W image obtained with
the {\it Hubble Space Telescope}.  By comparing the $K'$ versus
[F814W]$-K'$ color-magnitude diagram for the IC10 field with
theoretical isochrones, we find that the stellar population is best
represented by at least two bursts of star formation, one $\sim 10$
Myr ago and one significantly older ($150-500$ Myr). The young, blue
stars are centered around and concentrated in the vicinity of
[MAC92]~24. We suggest that this population represents the resolved
components of an OB association with a half-light radius of about 3
pc.  We resolve the W-R object [MAC92]~24 itself into at least six
blue stars.  Four of these components have near-IR colors and
luminosities that make them robust WN star candidates.
By matching the location of Carbon stars in the color-magnitude
diagram with those in the SMC we derive a distance modulus for IC~10
of about $24.5$ mag and a foreground reddening of $E(B-V) = 0.95$.  We
find a more precise distance by locating the tip of the giant branch
in the F814W, $H$, and $K'$ luminosity functions for IC 10.  We find a
weighted mean distance modulus $(m-M)_0 = 24.48\pm 0.08$. The
systematic error in this measurement, due to a possible difference in
the properties of the RGB populations in IC 10 and the SMC, is $\pm
0.16 $ mag.

\end{abstract}
\keywords{galaxies: individual (IC~10) -- galaxies: starburst --
  galaxies: star clusters} 

\section{Introduction}

Above the turbulence of the earth's atmosphere, and able to record
near-diffraction-limited images, instruments aboard the {\it Hubble
Space Telescope} ({\it HST}) have provided data that have driven the
field of high resolution imaging of starburst galaxies \citep[see,
e.g.,][and references therein]{2000astro-ph.0012546}.  While most of
the work with {\it HST} has been done at ultraviolet and visible
wavelengths, which are ideal for tracing hot, young stars, imaging and
spectroscopy in the near-infrared (near-IR) is invaluable when there
is substantial foreground or internal extinction or a large population
of red supergiants.  Because angular resolution at the diffraction
limit scales linearly with wavelength and inversely with telescope
diameter, the 2.5-m aperture of {\it HST} restricts its effectiveness
in the near-IR.  With the advent of adaptive optics (AO) facilities at
ground-based, 4-m and 8-m class telescopes, it is now possible to
acquire near-IR images with angular resolution superior to those from
{\it HST} and the NICMOS instrument.

AO observations require the presence of either an artificial or
natural guide star in the field of view. Such guide stars provide
input to the tip-tilt and wavefront sensors and, in the case of a
natural guide star, a straightforward means of determining the
point-spread function (PSF) of the system. Examples of extragalactic
studies using an AO system employing natural guide stars can be found
in \citet{2003PASP..115..635D} and references therein. Nevertheless,
care must taken when using a natural guide star in the field to avoid
saturation, persistence, and scattered light problems. Furthermore, in
general, extragalactic fields selected for stellar population studies
tend not to contain bright isolated stars.

With the deployment of laser guide star (LGS) techniques, the
requirement for a bright natural guide star within the isoplanatic
patch (a severe constraint when attempting to observe individual
extragalactic sources such as nearby starburst galaxies) can be
removed. The need for a tip-tilt guide star ($R \lesssim $ 17 mag.)
within the isokinetic angle ($\simeq 30''$) remains because, by
Fermat's principle, a monochromatic laser beacon cannot be used to
sense wavefront tilt. However, this constraint is much more easily
satisfied, even when observing sources out of the Galactic plane.

In this paper we report LGS AO near-IR images of the central region of
IC~10 obtained with the NIRC2 camera at the Keck Observatory.  IC~10
is a nearby irregular starburst galaxy, often characterized as a blue
compact dwarf \citep{2001A&A...370...34R}, located on the outskirts of
the Local Group. The physical properties of IC~10 are summarized in
the review by \citet{1998ARA&A..36..435M}.  IC~10 lies at low Galactic
latitude $(l=119.0^\circ,b=-3.3^\circ)$ and reliable values of the
foreground reddening and distance have been elusive, with estimates
ranging from 0.47 to 2.0 mag.\ for $E(B-V)$ and from 22 to more than
27 mag.\ for the distance modulus.  A summary of distance and
reddening estimates for IC 10 can be found in
\citet{1999ApJ...511..671S} and \citet{2004A&A...424..125D}.

IC~10 is remarkable for its high star formation rate, as evidenced by
its large number of H~II regions \citep{1990PASP..102...26H},
H$\alpha$ luminosity \citep{1998ARA&A..36..435M}, and far-IR
luminosity \citep{1994A&A...285...51M}. Normalized to its H$_2$
surface density, IC 10 has much higher rate of star formation than
most spirals or dwarfs
\citep{1998ARA&A..36..435M,2006ApJ...643..825L}.  IC~10 has a large
population of Wolf-Rayet (W-R) stars \citep{1992AJ....103.1159M,
1995AJ....109.2470M}, and the highest surface density of W-R stars
among all Local Group galaxies \citep{1995AJ....109.2470M}.
Furthermore, the ratio of carbon-type W-R's (WC stars) to
nitrogen-type W-R's (WN stars) in IC~10 is unusually high for its
metallicity of approximately 0.2-0.3 solar
\citep{1989ApJ...347..875S,1990ApJ...363..142G}.  These
characteristics suggest that IC~10 has experienced a brief but intense
galaxy-wide burst of star formation within the last $\sim 10$~Myr.

One of the most conspicuous sites of recent star formation in IC~10 is
the HII region [HL90] 111c \citep{1990PASP..102...26H}.  This region
also harbors the brightest putative W-R star in the galaxy,
[MAC92]~24, whose precise spectral type is uncertain.
\citet{2001A&A...370...34R} identified [MAC92]~24 as a WN star from
the broad He II $\lambda$ 4686 emission (and the lack of C IV $\lambda
4650$ emission) in its optical spectrum. Based on its He II $\lambda$
4686 equivalent width, \citet{2002ApJ...580L..35M} classified this
[MAC92]~24 as a WN star blended with another object.  However,
[MAC92]~24 was not recovered by \citet{2001A&A...366L...1R} in their
emission line survey of this galaxy, possibly because their narrow
band filters dedicated to W-R detection were contaminated by [O III]
$\lambda$ 5007 from [HL90] 111c.  \citet{2003A&A...404..483C}
subsequently noted that there were three closely-spaced sources,
[MAC92]~24-A, B, and C, all located within $1-2''$ of each other, that
might be contributing to the spectrum recorded by
\citet{2002ApJ...580L..35M}.  They also suggested that this object
might be an O3If WN star, a WN+OB binary, or a stellar cluster
containing a WN star.  The high stellar density in the vicinity of
[MAC92]~24 is apparent from the {\it HST}/WFPC2 images reported by
\citet{2001ApJ...559..225H} and \citet{2003A&A...404..483C}:
[MAC92]~24 lies at, or near the center of the young stellar cluster
[H2001] 4-1 visually identified by \citet{2001ApJ...559..225H}.  The
richness of this field, combined with the large foreground reddening,
makes IC~10 a highly attractive target for high resolution study with
near-IR laser-guide-star adaptive optics imaging techniques.

Our observations and data reductions are presented in \S2. The results
are presented in \S3 and discussed in \S4. A summary is given in \S5.

\vfill

\section{Observations and Data Reduction}
\subsection{Near-Infrared Data}
\label{SEC:ir-data}
Near-infrared $H$ and $K'$ images of IC~10, centered on [MAC92]~24 and
approximately 10 arc sec E and 18 arc sec S of [HL90] 111c, were
acquired with the NIRC2 camera on the Keck 2 telescope on 2005
November 10.  These observations were obtained using the laser-guide
star adaptive optics facility \citep{2006PASP..118..297W,
2006PASP..118..310V}.  The NIRC2 pupil stop was in the {\tt LARGEHEX}
position and the camera was operated in the narrow field mode, which
yields a scale of $0\farcs01$ per pixel and a field of view of
$10\farcs24 \times 10\farcs24 $.  The Nyquist frequencies with this
pupil stop on Keck 2 at $H$ and $K'$ are 65 and and 50 arc sec$^{-1}$,
respectively, and therefore the images are well sampled.

Observing conditions were photometric according to data from the Mauna
Kea all-sky monitor CONCAM\footnote{\url http://nightskylive.net/mk/},
and the natural seeing was measured to be $0\farcs3 $ full-width at
half maximum (FWHM) at $H$.  Analysis of the AO wavefront sensor
telemetry indicated that the Fried parameter was $r_0 \simeq 28$~cm
and the outer scale $L_0$ was $\sim 57$ m at 500 nm.  The airmass
during the observations was about 1.3. The Na I D$_2$ 589.0 nm laser
guide star, with an equivalent brightness of $R\simeq 11$ mag., was
projected at the center of the science camera field of view; a bright
star\footnote{USNO B1.0 1492-0009059, $I_{photo} = 13.59$ mag} located
due west $\sim 8''$ from the center of the field was used to perform
tip/tilt corrections.  The total exposure times were $1200$~s in $H$
(1.63 $\mu$m) and $1500$~s in $K'$ (2.12 $\mu$m), broken into four and
five dithered images, respectively. Each image in the dithering
sequence was separated from the previous one by $1\farcs0$.

\begin{figure*}
\plottwo{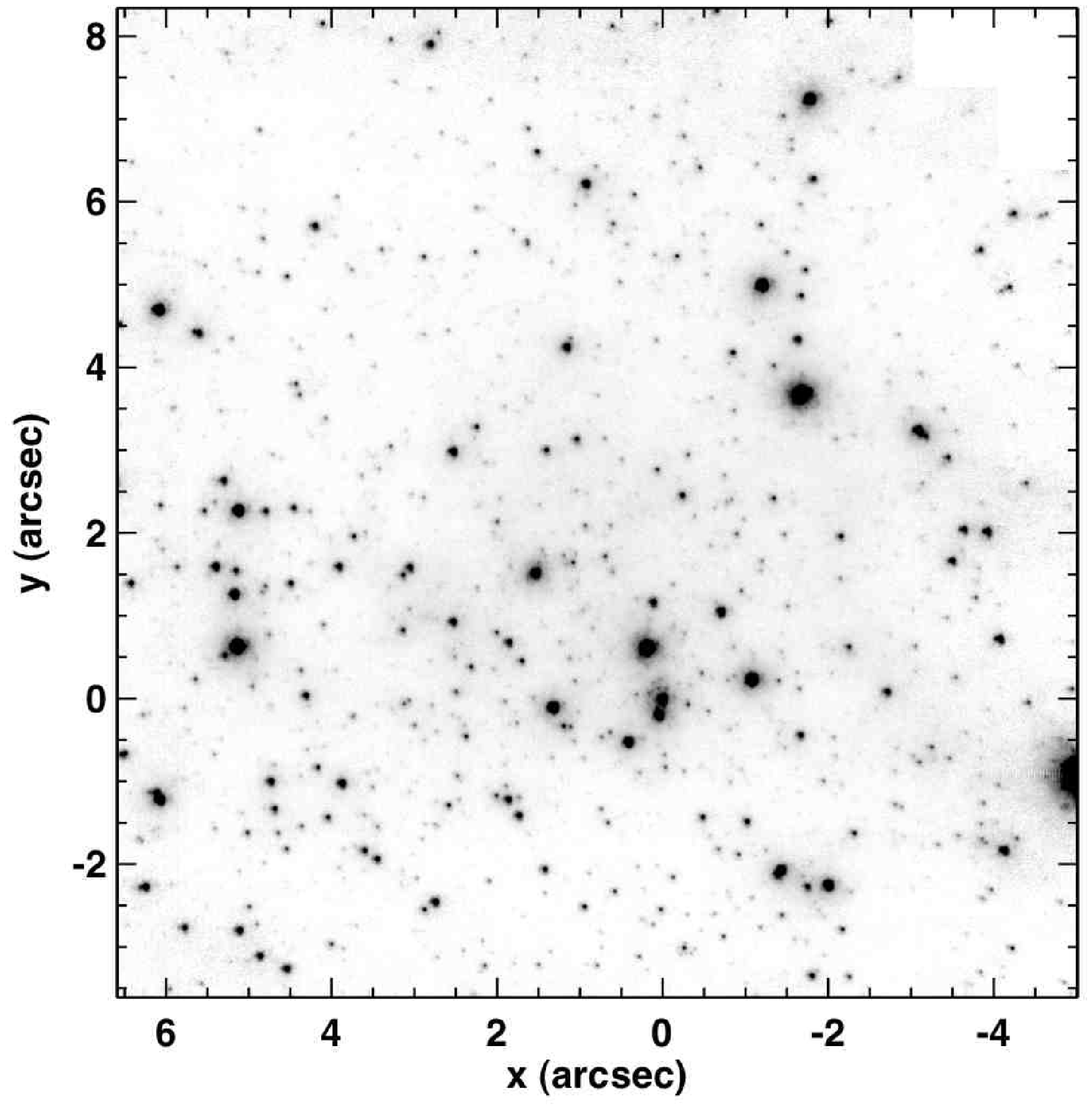}{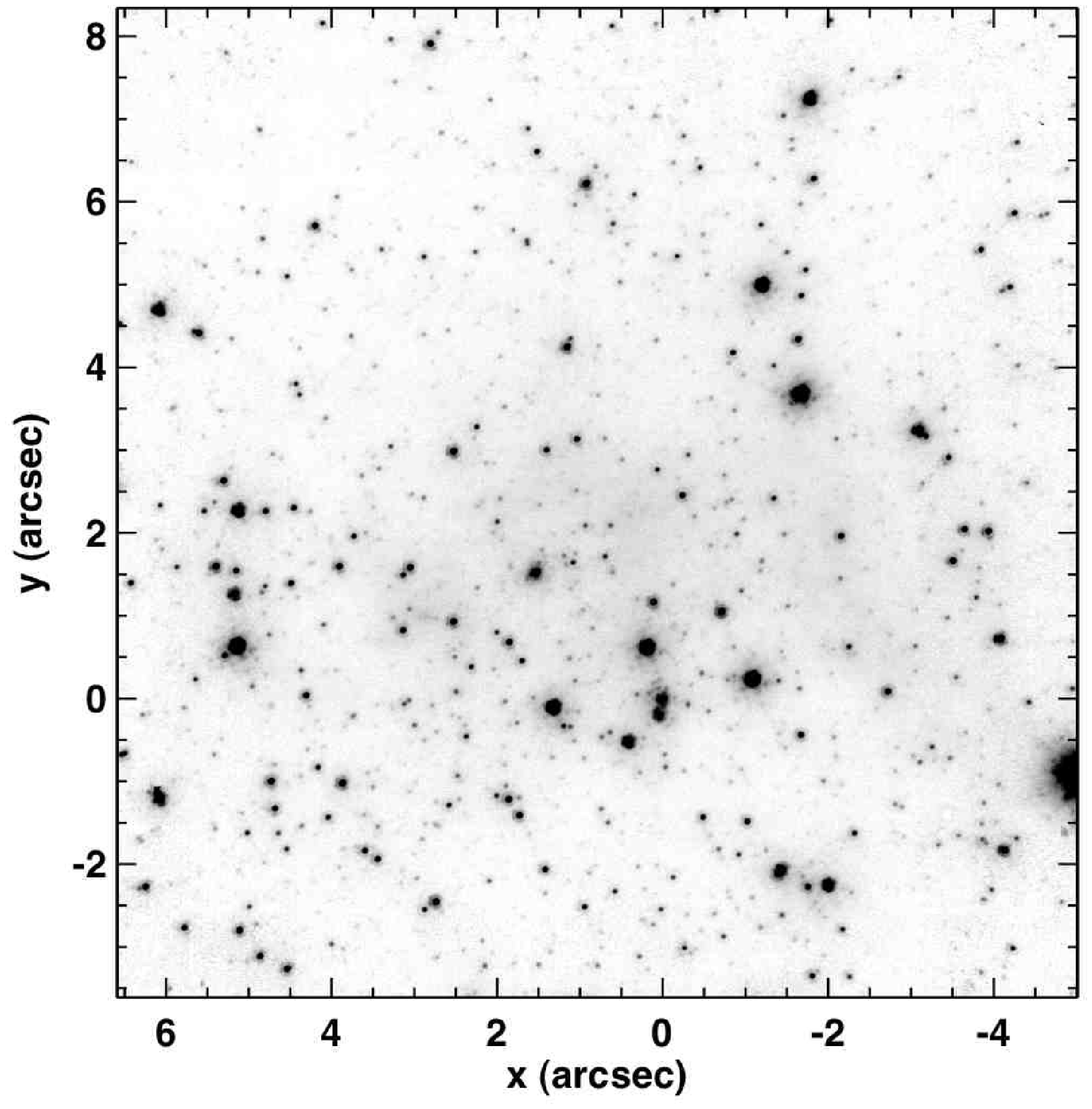}
\caption{Adaptive optics images of IC~10 obtained with the Keck 2
laser guide star system and the NIRC 2 facility camera. {\bf Left: }
$H$ band and {\bf Right:} $K'$ band mosaics about 20 arc sec SE of the
center of the HII region [HL90] 111c \citep{1990PASP..102...26H} and
include the W-R candidate [MAC]~24 \citep{1992AJ....103.1159M}.  The
images are near-diffraction limited with Strehl ratios of 18\% and
32\% at $H$ and $K'$, respectively.  The origin of the coordinate
system corresponds to the location of the northern component of the
W-R object [MAC92] 24-A at RA(J2000) $00^h$ $20^m$ $27^s.36$,
DEC(J2000) $+59^\circ$ $17'$ $37\farcs33$.  The orientation of the
images is conventional, with north up and east to the left.  The
stellar population is well resolved and faint nebular emission arcs
across the center of the images. }
\label{fig:HKmosaic} 
\end{figure*}

The raw images were reduced using our pipeline procedure for Lick and
Keck near-IR AO imaging data, which is currently maintained at
Berkeley by M.\ D.\ Perrin. The pipeline performs dark subtraction,
flat fielding, bad pixel repair, and stacking of images to create the
final mosaics.  Flat-fields were generated from exposures of the
interior of the dome obtained at the beginning of the night.  Dome
flats are preferable to twilight sky flats because they can be
acquired with high signal-to-noise.  The pupil illumination in the
Keck 2/AO/NIRC 2 system is well controlled and comparison of dome and
twilight sky flats shows only very small (negligible) differences,
presumably due to color and polarization differences between the two
sources.  As the data are well sampled, bad pixels in each $300$-s
exposure were repaired by replacing them with interpolated values
derived from their neighbors.  The dark-subtracted, flat-fielded
images were corrected for the known and previously mapped geometric
distortion introduced by the NIRC2 camera, registered to a common
coordinate frame, and combined.  The resulting mosaic in each filter
has a size of $12 \farcs 2 \times 12\farcs 2$, and the image size is
$0\farcs048$ FWHM at $H$ and $0\farcs051$ at $K'$.  The final images
are displayed in Figure \ref{fig:HKmosaic}.

Extracting photometry from the near-IR LGS AO images requires two
distinct steps: 1) PSF-fitting, which establishes the relative
brightness of stars; 2) analysis of the PSF to determine the
photometric curve of growth.  While conventional methods are
satisfactory to establish relative photometry, measuring the
photometric curve of growth needed to place AO measurements on an
absolute scale requires special techniques. In a crowded field,
relative photometry is best done using the narrow diffraction-limited
core. However, this core may contain as little as 5--10\% of the total
light.  The degree of AO correction is sensitive to the observing
conditions (seeing, wind speed, brightness of the wavefront reference,
etc.), and variations in these conditions causes variations in the
amount of energy in the core of the PSF relative to the uncorrected
seeing halo.  This ratio, known as the Strehl ratio ($SR$), is
exponentially sensitive to the variance of wavefront errors, and
Strehl variations can be large; for moderate correction
$\sigma_{SR}/SR \simeq 1$ \citep{2006ApJ...637..541F}.  As the
fractional photometric error is approximately equal to $\sigma_{SR}/SR
$, failure to measure and account for Strehl variations and the
corresponding change in the curve of growth can introduce order of
magnitude systematic errors in photometry.  Therefore, it is desirable
to estimate the PSF from the data themselves rather than from
observations of a PSF star at another time.  (This is a much more
straightforward task in the case of natural guide star AO
observations, when the guide star itself can be used to determine the
PSF and measure the photometric curve of growth. See, e.g.,
\citet{1999AJ....117.1297D, 2003PASP..115..635D}.)

\citet{2006ApJ...647...1517S} have shown that when the scene consists
of a rich star field the photometric curve of growth for
adaptively-corrected data can be extracted by modeling the power
spectrum of the image.  Modeling is performed in Fourier space and
employs physical descriptions of the optical transfer functions
(OTF) of the atmosphere, telescope, AO system, and science camera.
Although the Fourier power spectrum of the image of a single star is
proportional to $|{\rm OTF}|^2$, the power spectrum of a star field is
given by the product of $|{\rm OTF}|^2$ and the power spectrum of the
spatial source distribution. Therefore, both the imaging system and
the astronomical scene must be modeled.  By comparing Keck 2 and {\it
HST}/NICMOS data, \citet{2006ApJ...647...1517S} established that this
procedure reduces systematic errors in photometric measurements to the
few percent level.

\begin{figure}
\plotone{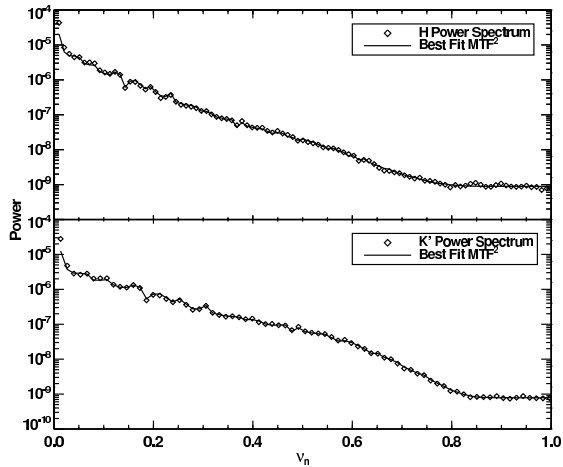}
\caption{Measured and best fit power spectra for the $H$ (top) and
$K'$ (bottom) band mosaics in Figure~\ref{fig:HKmosaic}.  Fitting the
optical transfer function (OTF) of the atmosphere, telescope, AO
system and science camera yields the Strehl ratio and allows the
photometric curve of growth to be evaluated
\citep{2006ApJ...647...1517S}.  Open diamonds show the one-dimensional
power spectrum extracted from the images.  Solid lines are the best
fit model power spectra computed from model system OTF multiplied by
the power spectrum of the source distribution.  Spatial frequency is
measured in normalized units relative to the Nyquist frequency at each
wavelength.  Comparison of the power at $\nu_n \simeq 0.6$ at $H$ and
$K'$ show that the correction at the longer wavelength is
superior. The corresponding Strehl ratios at $H$ and $K'$ are $18$\%
and $32$\%, respectively.  The first few wavenumbers are excluded from
the fits, as these are contaminated by flat fielding and sky
subtraction errors, as well as extended emission.  }
\label{fig:OTF_FIT} 
\end{figure}

\begin{figure}
\plotone{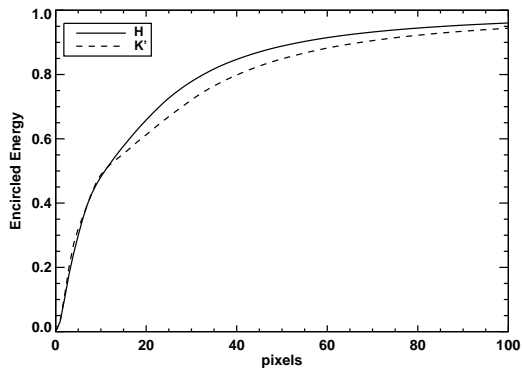}
\caption{Reconstructed curves of growth (encircled energy) as a function
of radius for the $H$ and $K'$ bands.}
\label{fig:EE_curves}
\end{figure}

We have applied the OTF-fitting algorithms described in
\citet{2006ApJ...647...1517S} to the current data set with no
modification.  The fits to the $H$ and $K'$-band power spectra are
shown in Figure \ref{fig:OTF_FIT}. Note that the lowest wavenumbers
are excluded from each fit in order to mitigate the effects of
possible contamination of the power spectra by flat-fielding and sky
subtraction errors, and extended (nebular) emission. 
Based on the uncertainties in the reconstruction of the
photometric curves of growth (encircled energy curves), which are
shown in Figure \ref{fig:EE_curves}, we estimate that the systematic
errors in the photometry are $\leq 0.04$ mag.\ in both bands.  The
reconstructed PSF implies that the Strehl ratios of the final mosaics
are $18$\% in $H$ and $32$\% in $K'$.

We investigated PSF variations across the field of view by segmenting
the final mosaics into a number of separate sub-regions, performing
the analysis described above on each sub-region separately, and
constructing the encircled energy curves for each region.  Comparison
of the results show variations of the aperture corrections of less
than 5\%.  Furthermore, we find no systematic variation of the
aperture correction with position in the field.  These results agree
with both visual inspection of the images, which reveals no
significant astigmatism at the edges of the field, and other
measurements of anisoplanatism in AO images obtained under
typical seeing conditions at Mauna Kea
\citep[e.g.,][]{1999AJ....117.1297D,2003PASP..115..635D}.  Hence, we
conclude that the isoplanatic angle was large for these observations
and we neglected any PSF variations over the field of view in our
analysis.

We established the photometric zero point using observations of the
UKIRT near-IR standard star FS6 \citep{2001MNRAS.325..563H}, obtained
on the same night.  We did not observe enough standards stars to
determine the atmospheric extinction coefficient from our data.
However, the $H$ and $K'$ filters in NIRC2 are similar to those in use
at the IRTF, as they were produced as part of the Mauna Kea filter
consortium \citep[see][]{2002PASP..114..180T}.  Therefore, we have
used the nominal atmospheric extinction coefficients of 0.059 mag.\
airmass$^{-1}$ and 0.088 mag.\ airmass$^{-1}$ for $H$ and $K'$
respectively given on the IRTF web site\footnote{
\url{http://irtfweb.ifa.hawaii.edu/$\sim$nsfcam/hist/backgrounds.html}; \\
see also \\
\url{http://www.jach.hawaii.edu/UKIRT/astronomy/calib/phot\_cal/cam\_zp.html}
} to make the small corrections necessary to account for the small
difference in airmass between our science target and the standard
star.

We performed relative photometry by constructing an empirical PSF with
a radius of $0\farcs19$ from seven bright and relatively isolated
stars in the field using DAOPHOT \citep{1987PASP...99..191S}.  We then
used the crowded field photometry package StarFinder \citep{diolaiti}
to perform source detection and PSF fitting.  Photometry resulting
from PSF fitting returns the relative brightness and locations of the
stars in the field of view, and therefore provides the information
necessary to estimate the shape of the power spectrum of the source
distribution.

\begin{figure}
\plotone{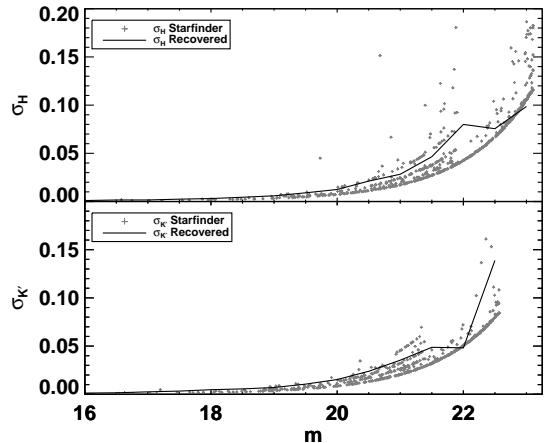}
\caption{ Two estimates of the internal errors in the $H$ and $K'$
band photometry. Grey points are photometric errors returned by {\tt
StarFinder} for individual stars.  The solid lines are the photometric
errors derived from an artificial star test.  The {\tt StarFinder}
errors vary with field position because the exposure time is
non-uniform across the mosaic. The solid lines from the artificial
star experiment represent an average over the field of view.  }
\label{fig:HKerrors} 
\end{figure}

We identified 661 stellar sources in the $H$ image and 585 sources in
the $K'$ image.  For the brightest few stars, speckles in the PSF halo
that lie outside the $0\farcs19$ empirical PSF radius were erroneously
identified as sources.  As the location of speckles varies with
wavelength, these spurious detections were easily identified and
eliminated by hand.  A variance map for each mosaic was constructed
using the formulae given by \citet{2004PASP..116..352V} and supplied
to StarFinder to calculate the internal photometric errors.  The
photometric errors returned by StarFinder are shown in Figure
\ref{fig:HKerrors}.  Several discrete tracks are evident in this
figure because the total exposure time, and hence the signal-to-noise
ratio, is non-uniform across the mosaics.  For example, the total
exposure time for the central area of the five-point dither pattern
was five times larger than that at the edges.

\begin{figure}
\plotone{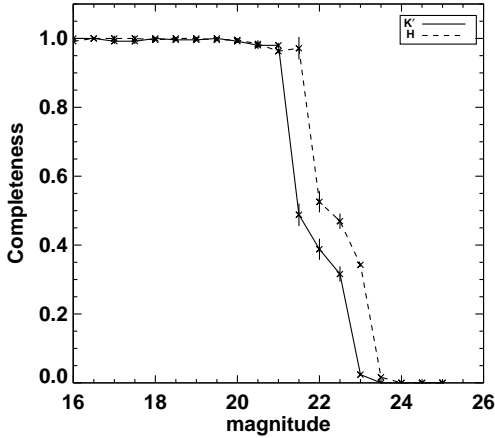}
\caption{
Photometric completeness of the $H$ and $K'$ band photometry as a
function of limiting magnitude, derived from the artificial star
tests.
}
\label{fig:HKcomplete} 
\end{figure}

Photometric completeness limits were determined by inserting
artificial stars of known magnitude into each mosaic and attempting to
recover them. We performed this procedure for magnitudes between
$16.0$ and $25.0$ in steps of $0.5$ mag. To accumulate good statistics
without substantially increasing the crowding in the images, we
restricted the number of stars inserted to $50$ and repeated the
procedure five times for each magnitude bin.  Figure
\ref{fig:HKcomplete} presents the resulting completeness as a function
of magnitude.  The limiting magnitudes are 22.3 and 21.5 in $H$ and
$K'$, respectively, for a completeness threshold of 50\% recovery.  In
order to test the photometric errors presented in Figure
\ref{fig:HKerrors}, we measured the difference between the input
magnitudes and the recovered magnitudes for each artificial star. The
width of the best-fit Gaussian to these residuals as a function of
input magnitude is shown as the solid line in Figure
\ref{fig:HKerrors}.  The agreement between the two error estimates is
satisfactory.

We note that our near-IR photometry is substantially deeper and more
accurate than that presented by \citet{2000A&A...363..130B}. Our
completeness limits are $\sim 5$ and $\sim 3$ mags deeper in $H$ and
$K'$, respectively, and at any given magnitude, our internal errors
are a factor of $\sim 5$ smaller. Of course, our field of view is
considerably smaller ($0.04$ arcmin$^2$ vs.\ $13$ arcmin$^2$).

\subsection{HST Data}
\label{SEC:hst-data}

Optical imaging was obtained with the Wide Field Camera (WFC) of the
Advanced Camera for Surveys (ACS) \citep{2003SPIE.4854...81F}.  Three
F814W-band images were acquired as part of program GI-9683 (PI: Bauer)
and were retrieved from the {\it HST} archives\footnote{ACS/WFC F435W
and F606W images are also available in the archives as part of this
program. We chose to use only the F814W data because the shorter
wavelength images suffer from greater undersampling and greater
sensitivity to foreground and internal extinction.}.  The total
exposure time was 1080 s.  We used the {$\tt \_flt$} (calibrated
individual exposure) and {$\tt \_crj$} (calibrated, cosmic
ray-split-combined image) files processed with the standard {\it
HST/ACS} pipeline, which included corrections for the bias level, dark
current, flat fielding, flux calibration and (for the {$\tt \_crj$}
files) cosmic ray rejection and image combining. The mean ACS/WFC
pixel scale is approximately $0\farcs05$.  Since the Nyquist frequency
at this wavelength on {\it HST} is 30 arc sec$^{-1}$, these data are
undersampled.

\begin{figure}
\plotone{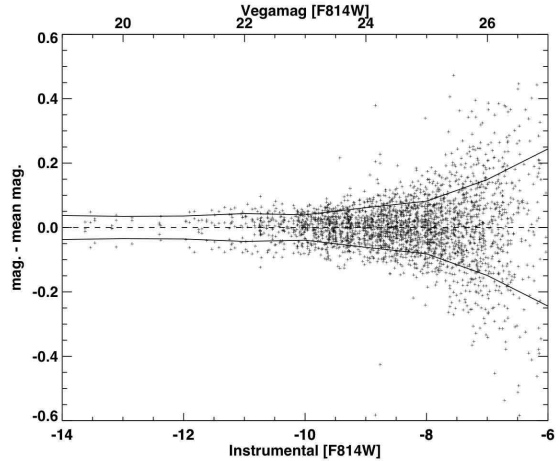}
\caption{ Internal errors in the F814W photometry were estimated by
comparing photometry from the three individual {\tt \_flt} 
images. This plot shows the residuals
between the {\tt \_flt} magnitudes and the mean 
magnitude.  The solid line shows the standard deviation of the
residuals in 1 mag.\ wide bins.  For [F814W] $<$ 23 mag.\ the rms
internal error is approximately 0.04 mag.  }
\label{fig:I_errors} 
\end{figure}

Photometry was performed using the software described by
\citet{anderson06}. This code identifies sources on the images above a
sky threshold, generates a spatially variable PSF for the specific
filter, and fits it to each object detected on the image to yield a
distortion-corrected instrumental magnitude and position. The PSF is
normalized to have a flux of unity within a radius of 10 pixels, or
$0\farcs5$. Although this code was developed for use on the {$\tt
\_flt$} images, we found only a small systematic difference (0.025
mag.)  between the instrumental magnitudes derived from the {$\tt
\_flt$} images and those derived from the {$\tt \_crj$} image.  The
latter has the advantage of eliminating many spurious sources due to
detector artifacts and cosmic rays.  Furthermore, this difference is
well within the RMS error found by examining the scatter of the
photometry derived from the three individual {\tt \_flt} images about
the mean (Fig. \ref{fig:I_errors}).  This comparison reveals that the
uncertainties are $<0.04$ mag.\ for objects brighter than ${\rm
[F814W]}=23.0$ mag.

Instrumental magnitudes were converted to a Vega-based system by
adopting a zero-point of 25.501 mag.  We shall refer to these as
[F814W] magnitudes.  We have not transformed these values to a
ground-based system \citep[e.g.,][]{1988PASP..100.1134B}.  Restricting
the analysis to stars within the NIRC2 field, we find $\sim 870$
sources brighter than ${\rm [F814W]}=26.5$ mag.  Our estimated 50\%
completeness limit is between 25 and 25.5 mag, for which we find $\sim
690$ sources.

Stars in the NIRC2 field were matched with stars in the ACS field
using a nearest-neighbor algorithm.  Detector coordinates of sources
in the geometrically-distorted ACS frame were transformed into sky
coordinates using the
PyRAF\footnote{\url{http://www.stsci.edu/resources/software\_hardware/pyraf}}
utility {\tt tran}, and back into NIRC2 detector coordinates.  As the
pixel scale of the ACS/WFC is $5$ times coarser than that of the NIRC2
narrow camera and pixel mismatches may be large, we selected sources
that had common centroids within a $7$ NIRC2 pixel radius (70 mas).
To further prevent false associations in the automatically generated
catalog, we restricted our matches to stars separated from one another
by at least this distance in both individual catalogs.  In the crowded
region around [MAC92]~24 the threshold for matching was lowered and
each potential match was inspected by hand before being added to the
catalog. This procedure resulted in [F814W]$-K'$ colors for 380 stars.

\section{Results}

\subsection{Three-Color Image}
\label{sec:three-color-image}

\begin{figure*}
\plotone{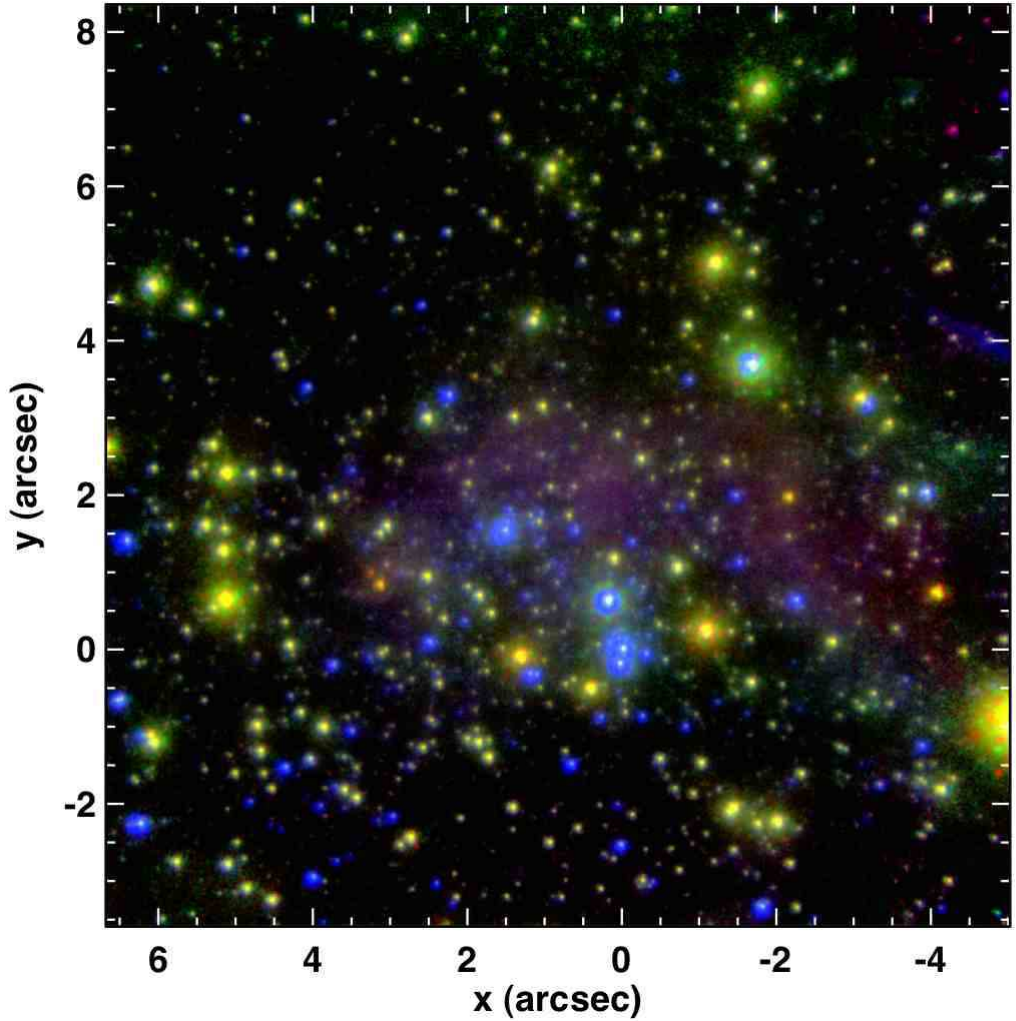}
\caption{Three-color $IHK'$ composite of the [HL90] 111c HII region of
IC~10.  The coordinate system is centered on the W-R 
candidate [MAC92]~24-A/N.  
North is up and east is to the left.  The near-IR $H$ and
$K'$ images are laser guide star adaptive optics data from the Keck 2
telescope obtained with NIRC2, and the $I$-band data is F814W from the
Advanced Camera for Surveys on the {\it Hubble Space Telescope}.  This
image was constructed using the algorithm of
\citet{2004PASP..116..133L}, which ensures that an object with a
specified astronomical color has a unique color in the RGB image.
Distinct populations of red and blue stars are evident. The blue stars
comprise main sequence stars and blue supergiants; the red stars are
predominantly red giants. The brightest red stars are AGB stars. The
reddest stars are carbon stars and evolved AGB stars with
thick circumstellar envelopes. 
Consult the color-magnitude diagram
(Fig. \ref{fig:CMD}) and the finding chart
(Fig. \ref{fig:kprimefindingchart}) to identify specific objects.
Diffuse light traces nebular emission from the 111c HII region.  }
\label{fig:IHK} 
\end{figure*}

\begin{figure}
\plotone{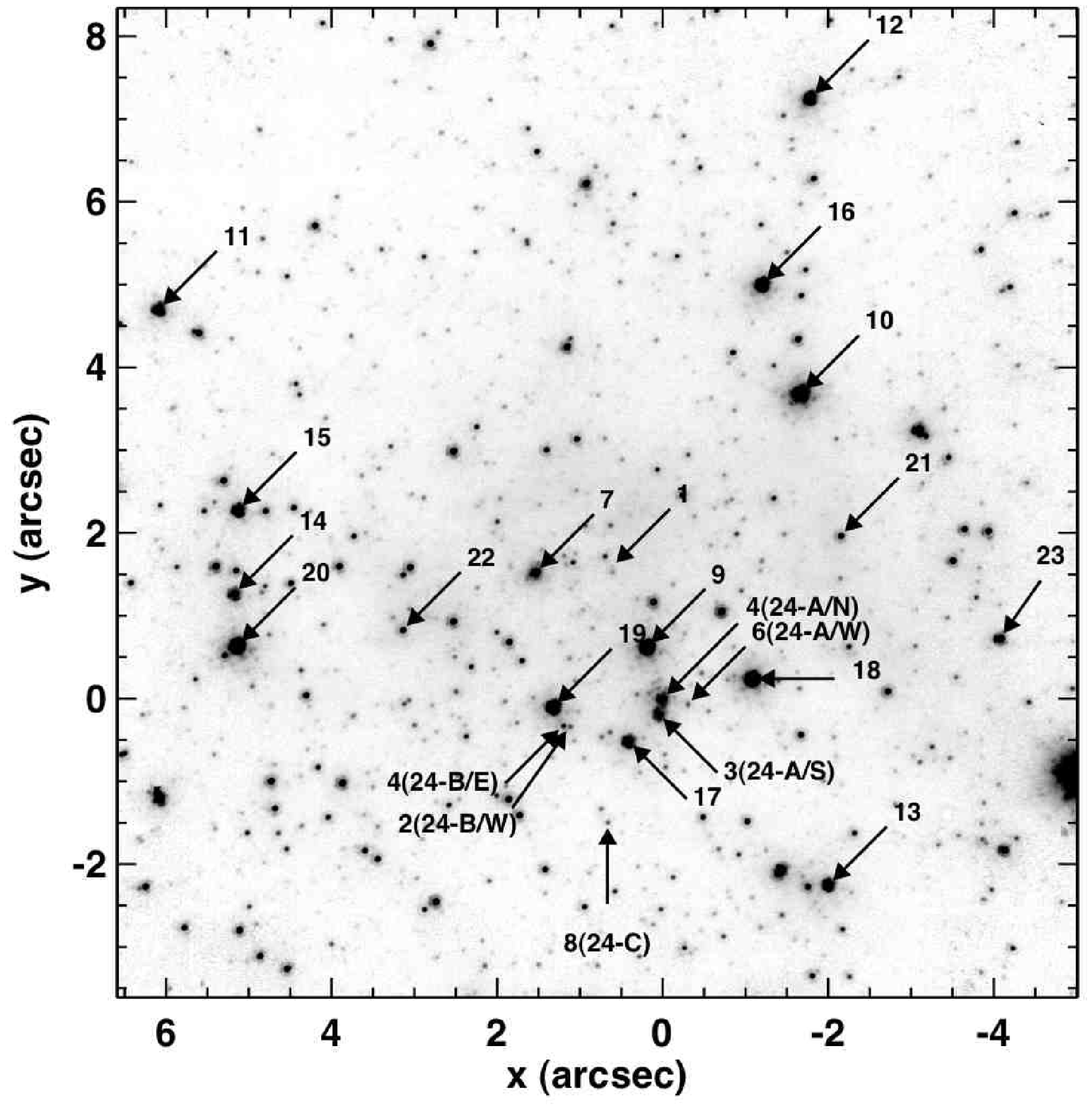}
\caption{$K'$ finding chart with scale and orientation matching
Fig. \ref{fig:IHK}.  The components of [MAC92] 24-A and B are labeled
WR24-A and WR24-B, respectively.  The object number can be used to
find the stars' location on the $K'$/ [F814W]$-K'$ color-magnitude
diagram (Fig. \ref{fig:CMD}). Note that the object numbering scheme
runs from left to right in Fig. \ref{fig:CMD}. Thus, object number 1
is the bluest object and object number 22 is the reddest.  }
\label{fig:kprimefindingchart} 
\end{figure}

A three-color image of the IC~10 field, combining the F814W, $H$, and
$K'$ data and produced using the algorithm described by
\citet{2004PASP..116..133L}, is shown in Figure \ref{fig:IHK}.  The
\citet{2004PASP..116..133L} algorithm ensures that an object with a
specified astronomical color has a unique color in the RGB image.  The
color image draws attention to the spatial distribution of the red and
blue sequences that are evident in the color-magnitude diagram (CMD;
see Fig. \ref{fig:CMD}) and serves as a useful aid in the
identification and discussion of individual sources.  In Figure
\ref{fig:kprimefindingchart} we show an annotated version of the $K'$
image of this field.  We have labeled the brightest, bluest and
reddest objects in the image, as well as other sources whose locations
in the color-magnitude diagram are noteworthy.  Photometry for these
objects is listed in Table \ref{TAB:phottable}.

\subsection{The Color-Magnitude Diagram}
\label{sec:CMD}

\begin{figure}
\plotone{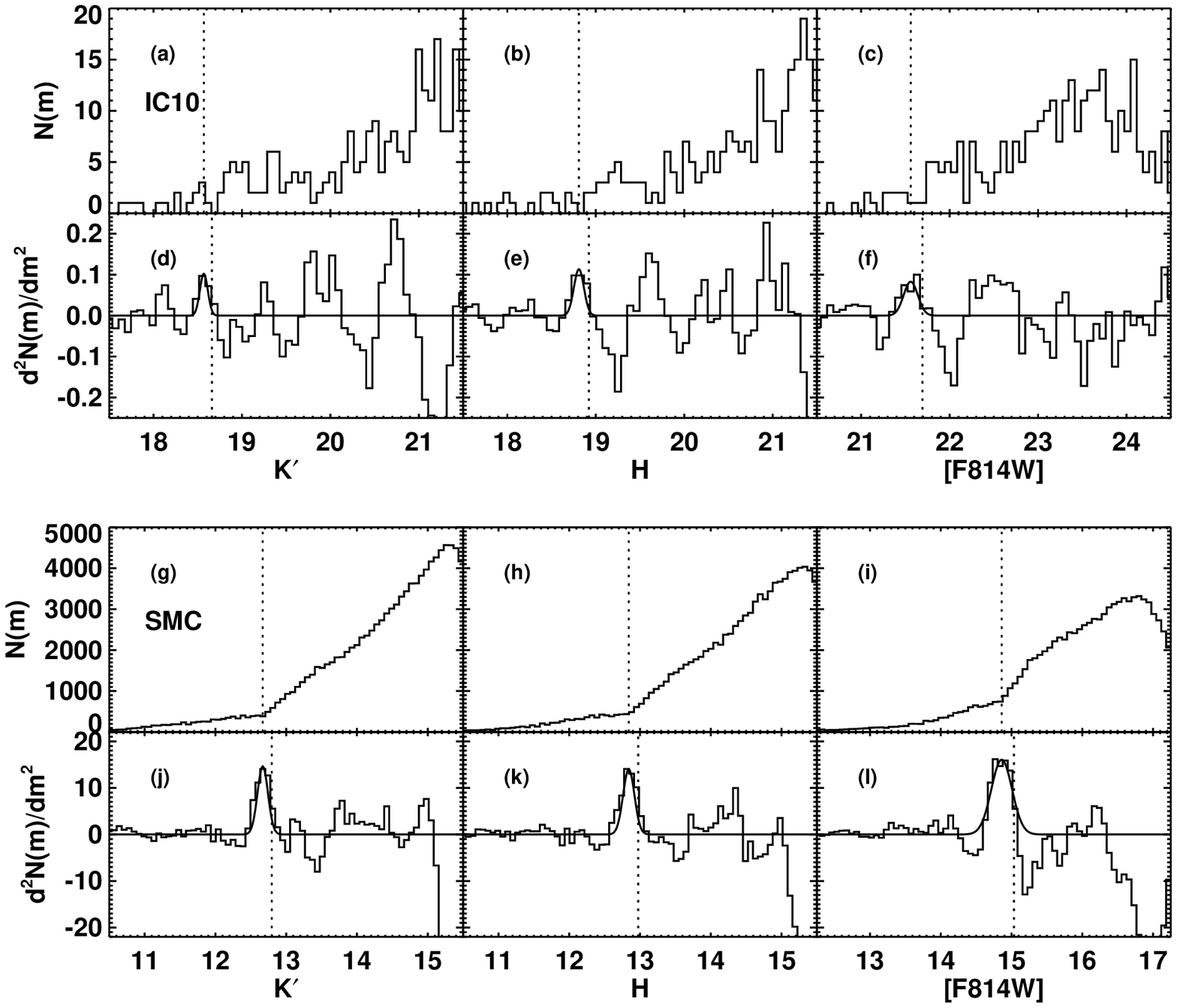}
\caption{Top six (a--f) panels show luminosity functions in $K'$, $H$,
and [F814W] (a--c, respectively) their corresponding second
derivatives (d--f) for IC~10.  The bottom six panels (g--l) show the
same plots for the SMC.  The second derivative (lower row of each
set), which is used to locate the tip of the red giant branch, is
computed using according to the recipe of \citet{2000A&A...359..601C}
including smoothing with a Savitzky-Golay filter and a correction that
accounts for the systematic offset between peak in the second
derivative and the location of the TRGB.  The displacement between the
vertical lines in the upper and lower panels indicates the magnitude
of this correction.  The light grey line is the best fit Gaussian to
the second derivative peak.  }
\label{fig:LF} 
\end{figure}

\begin{figure}
\plotone{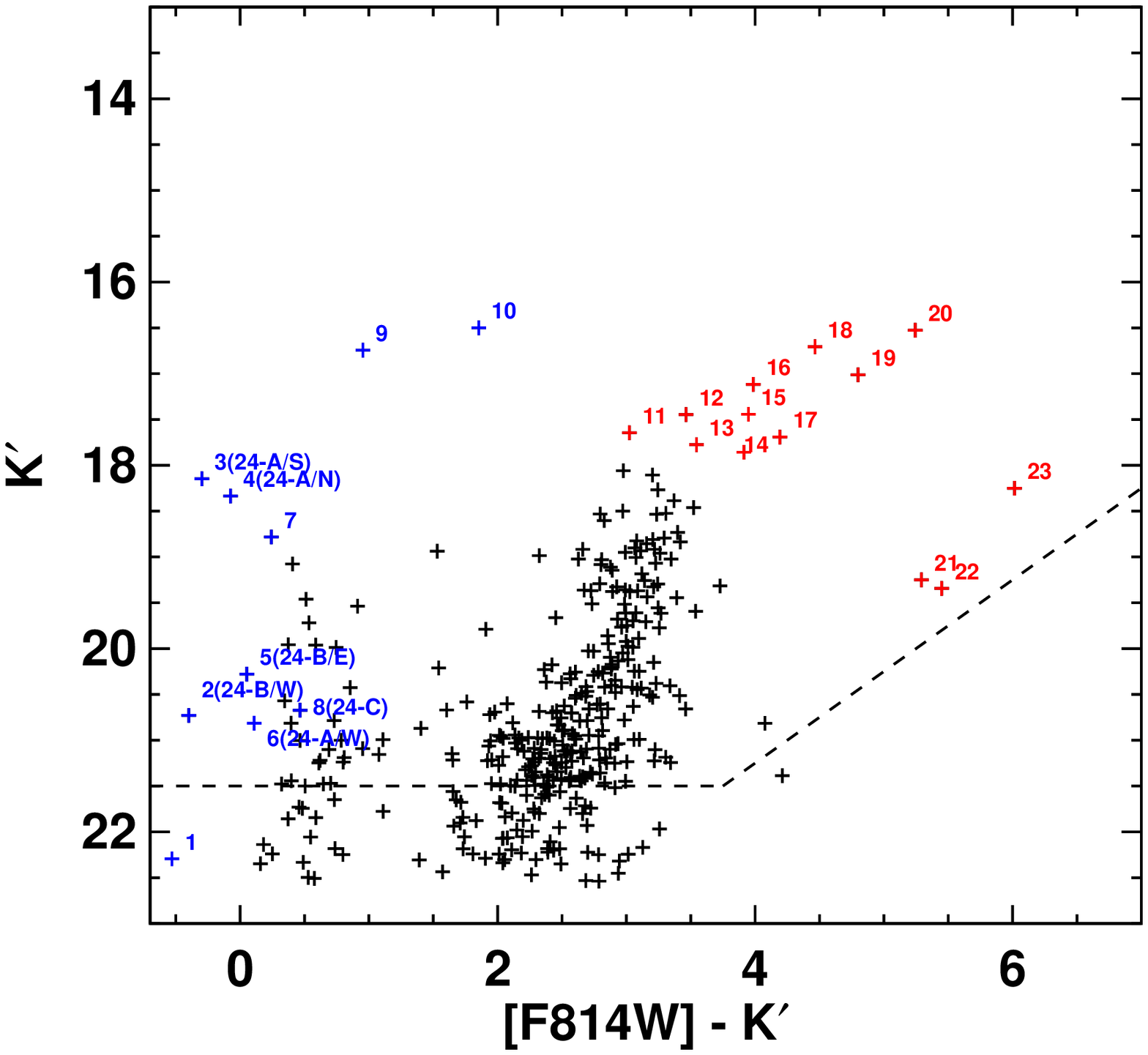}
\caption{$K'$ vs $I-K'$ color-magnitude diagram for the NIRC2 field.
Consult Fig. \ref{fig:kprimefindingchart} to identify the location of
the numbered objects. The dashed line is the 50\% completeness limit.
}
\label{fig:CMD} 
\end{figure}

From our photometric catalog we generated the [F814W], $H$, and $K'$
luminosity functions as well as the $K'$ versus [F814W]$-K'$ CMD. The
former are shown in Figure~\ref{fig:LF} while the latter appears in
Figure~\ref{fig:CMD}.  Two distinct sequences are populated in the
CMD: a vertical band of blue stars with [F814W]$-K' \simeq 0.5$ mag.,
which are most likely reddened main sequence objects, and a
well-populated plume of red stars with ${\rm [F814W]}-K'$ in the range
2--3.5 mag.  In addition there are several very red objects with
[F814W]$-K' > 4$ mag.

We estimated the contamination of foreground Galactic stars in our
CMDs using the Galactic models of \citet{1994AJ....107..582C}.
Although IC~10 is at low Galactic latitude, foreground contamination
is not severe in such a small field.  In $H$ band, we find that we
expect about 0.01 stars arcsec$^{-2}$ brighter than 18 mag., 0.025
stars arcsec$^{-2}$ brighter than 20 mag., and 0.048 stars
arcsec$^{-2}$ brighter than 22 mag.  These stellar surface densities
correspond to 1, 4, and 7 stars, respectively, in our NIRC2
field. Similar values are found for the $ K'$ band.  Thus, above our
50\% completeness limit we suffer about 1\% contamination.  The
contaminating stars are expected to be main sequence dwarfs, and
therefore should have moderately red colors with ${\rm [F814W]}-K'$ in
the range 0.8--2.2 mag.  We do not expect that any of the blue stars
($ {\rm [F814W]}-K' \simeq 0.5$ mag.)  or any of the extremely red
stars ($ {\rm [F814W]}-K' > 4$ mag.)  are foreground objects.  Two
bright ($K' \simeq 17$ mag.)  stars, labeled 9 and 10 in Figure
\ref{fig:CMD}, have the colors of foreground dwarfs. However, the
close spatial association of star 9 with [MAC92]~24 suggests that it
is a reddened supergiant in IC~10, with an initial main sequence mass
of $\simeq 25 M_\odot$. The status of star 10 is uncertain.

We also investigated possible contamination by background galaxies.
Based on the galaxy counts given by \citet{2001ApJ...560..566K} and
\citet{2001MNRAS...325..550D} and the size of our fields, we would
expect to find less than one background galaxy in our $K'$ band image
down to our 50\% completeness limit of 21.5 mag. This agrees with our
finding that most of the objects in our image are point-like (see
below).

\begin{figure*}
\plotone{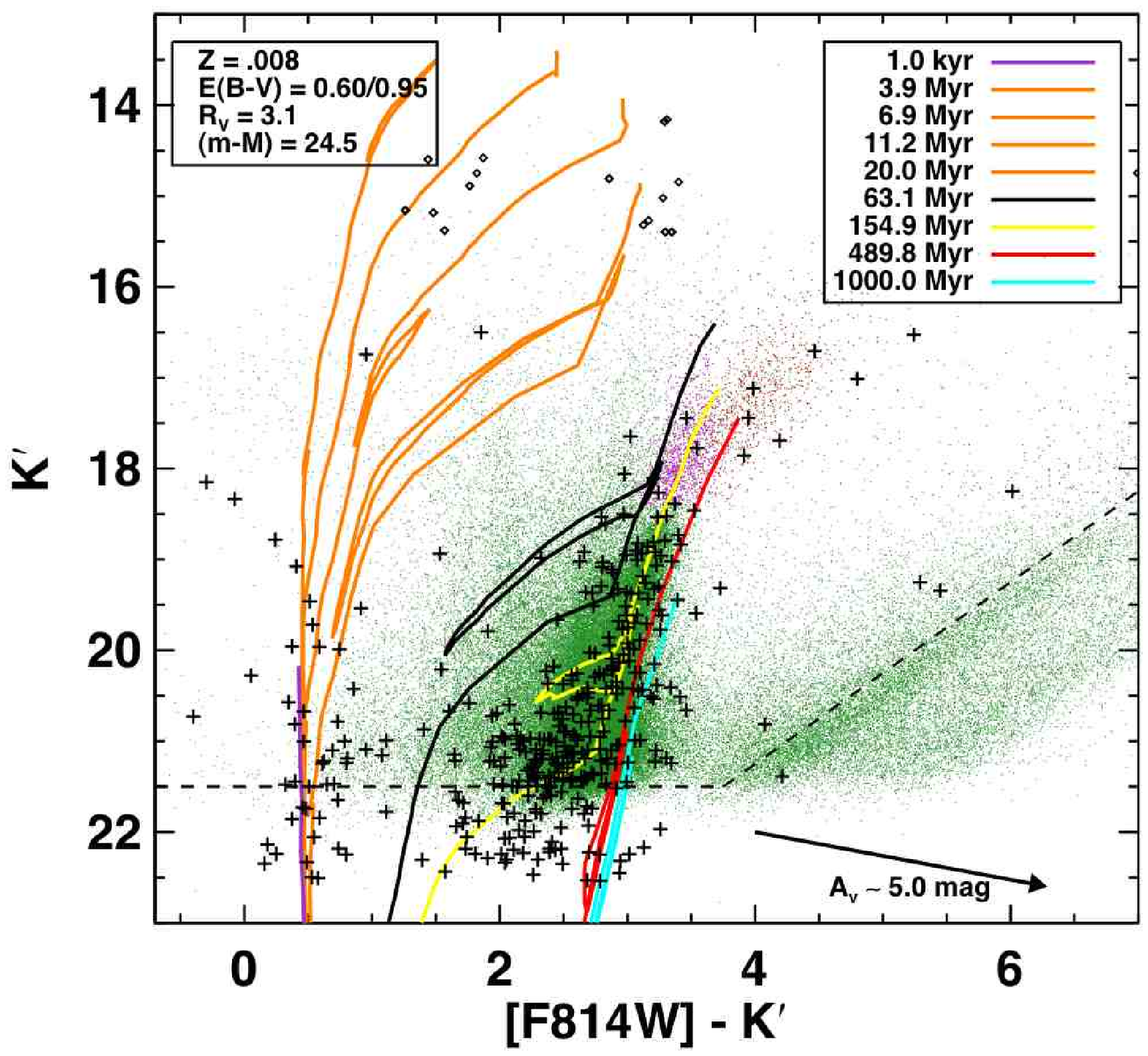}
\caption{ $K'$ versus [F814W]$-K'$ color-magnitude diagram for IC 10.
The solid lines denote the theoretical $Z = 0.008$ isochrones from
Lejeune \& Schaerer (2001). The initial mass function terminates at an
upper mass of 120 $M_\odot$.
Colored dots are
stars from the SMC photometric
catalog of \citet{2002AJ....123..855Z}, including O-rich (magenta) and
C-rich (red) AGB stars.  The rest of the SMC population, including the
red giants and very red, dust enshrouded objects are plotted as green.
Both the isochrones and the SMC data have been reddened and shifted in
magnitude, using the values given in the upper left legend to match
the IC~10 data The isochrones corresponding to ages $<$ 20
Myr have been reddened by $E(B-V) = 0.60$ mag., while the isochrones
for older ages have been reddened by $E(B-V) = 0.95$ mag. The SMC data
have been reddened by $E(B-V) = 0.95$ mag.
Open diamonds are
spectroscopically-confirmed
Magellanic Cloud red supergiants from \citet{2002AJ....123..855Z}.
According to artificial star experiments the IC 10 photometry is 50\%
complete at the dashed black line.  }
\label{fig:CMD_compare} 
\end{figure*}

The metallicity of the SMC is close to that of IC 10, and therefore
this nearby system forms a natural template against which IC~10 can be
compared.  We used the 2MASS and DENIS photometric catalogs for the
SMC provided by \citet{2002AJ....123..855Z} for this task. We
transformed the 2MASS magnitudes to the NIRC2/MKO system using the
relations given in \citet{2001AJ....121.2851C} along with an estimate
of the conversion between $K$ and $K'$ filter magnitudes as a function
of $H-K$ color\footnote{$(K' - K) \simeq 0.11 (H-K)$. This relation
was derived using the method of Wainscoat \& Cowie (1992), based on
the central wavelengths of our filters}.  We used the relations given
by \citet{2005PASP..117.1049S} to convert Cousins $I$ band magnitudes
given by \citet{2002AJ....123..855Z} to HST/ACS/WFC [F814W].
\citet{2006A&A...452..195C} identify C-rich and O-rich asymptotic
giant branch (AGB) stars in the SMC based upon their $J$ and $K_s$
photometry in the DENIS and 2MASS catalogs.  Using their criteria for
partitioning the $K_s$ versus $J-K_s$ CMD into regions of distinct
stellar types, we identified AGB stars in the $K_s$ versus $J-K_s$ CMD
of \citet{2002AJ....123..855Z} and carried the divisions over to the
$K'$ versus ${\rm [F814W]}-K'$ CMD.  The O-rich and C-rich AGB stars
in the SMC sample are plotted in Figure~\ref{fig:CMD_compare} as
magenta and red points, respectively.  The rest of the SMC stars,
including the red giant branch, are plotted as green points.  We
assumed a distance modulus to the SMC of $18.95\pm0.05$ mag.\ and a
reddening of $E(B-V) = 0.15\pm0.06$ mag.\ for consistency with
\citet[][]{2000A&A...359..601C}.

To compare the SMC and IC~10 CMDs directly (in the same figure), it is
necessary to adopt a foreground reddening and a distance modulus for
IC~10.  After experimenting with a range of values found in the
literature \citep[see][]{2004A&A...424..125D}, we found that a good
match between the SMC and the IC~10 distributions could be achieved
with a foreground reddening of $E(B-V)=0.95\pm0.15$ mag.\ and a
distance modulus of 24.5 mag. The reddening is determined primarily by
matching the ${\rm [F814W]}-K'$ color of the red giant branch
\citep[cf.][]{1999ApJ...511..671S}.  
If we focus on the less abundant blue stars,
and attempt to match their location in the CMD with that of the very
top of the main sequence in the SMC distribution, we find that a
somewhat smaller reddening value of $E(B-V) \sim 0.65$ mag.\  is
preferred. This estimate has a large uncertainty, however, given the
small number of blue main-sequence stars above the completeness limits
in both the IC 10 and SMC samples.  We note that IC~10 has abundant
molecular clouds \citep{2006ApJ...643..825L} and there is ample
evidence for substantial internal reddening. The differential
extinction between the blue and red stars may indicate that the blue
stars lie on the front face of IC~10, in a region that has been
cleared of gas and dust by the O-star winds and supernovae associated
with the star formation activity evident in the vicinity of [HL90]
111c.

The comparison with the SMC indicates that the red population in IC 10
comprises both red giants and AGB stars. There are two branches of
very red (${\rm [F814W]} - K' > 4$ mag.) stars found in our IC~10
field.  C-rich AGB stars comprise the bright branch.  Highly evolved
AGB stars with optically thick circumstellar dust envelopes populate
the dim branch.  Additional red stars between $K'$ = 20 and 22 mag.\
are undoubtedly present, but the lower right corner of the CMD is
unpopulated due to incompleteness in the F814W measurements.

We have identified red supergiants (RSGs) in the
\citet{2002AJ....123..855Z} catalog of SMC stars and plotted them on
Figure~\ref{fig:CMD_compare}.  We began with the list of
spectroscopically-confirmed Magellanic Clouds RSGs from
\citet{2003AJ....126.2867M}.  Cross-correlation of the coordinates and
$B$ and $V$ magnitudes of these RSGs with objects in the
\citet{2002AJ....123..855Z} catalog yielded their location in the
[F814W] versus [F814W]-$K'$ CMD. This exercise indicates that there
are no unambiguous RSGs within our field of view, except for perhaps
the bright star located on the southwest edge of the frame for which
we did not attempt to perform photometry.

In Figure~\ref{fig:CMD_compare} we have overlaid theoretical stellar
isochrones for $Z = 0.008$ from \citet{2001A&A...366..538L} on our
observed CMD.  The oxygen abundance in IC 10 in HII regions
corresponds to $Z \simeq 0.005$ ($0.25 Z_\odot$)
\citep{1989ApJ...347..875S, 1990ApJ...363..142G}.  Thus, one would
expect that $Z = 0.004$ tracks of \citet{2001A&A...366..538L} would be
most suitable for such a comparison. However, as found by
\citet{2001ApJ...559..225H}, the more metal-rich $Z = 0.008$
isochrones extend farther to the red than the $Z = 0.004$ isochrones
do and so encompass the observed location of red supergiants.  These
tracks also better match the observed integrated colors of star
clusters in IC 10 \citep{2001ApJ...559..225H}.  We converted the WFPC2
F814W magnitudes given by \citet{2001A&A...366..538L} to the ACS/WFC
F814W band using the prescription given by
\citet{2005PASP..117.1049S}.  The \citet{1988PASP..100.1134B} $H$ and
$K$ band magnitudes given by \citet{2001A&A...366..538L} were
converted to the NIRC2/MKO system using the transformations given by
\citet{2001AJ....121.2851C}, along with the aforementioned correction
between the $K$ an $K'$ filter magnitudes as a function of $H-K$
color. We shifted the isochrones to correspond to a distance modulus
of 24.5. We also reddened the isochrones, with the adopted reddening
value dependent on the corresponding age of each isochrone. This was
done in order to account for the differing reddening values found
above for the red and blue stars. For ages of $\leq 20$ Myr, we found
that a reddening value of $E(B-V) = 0.60$ mag.\ yields a good match
between the theoretical main sequence given by the models and the
location of the blue stars in the IC 10 CMD.  This reddening value is
somewhat lower than the foreground extinction of $E(B-V) = 0.8\pm 0.1$
mag.\ recommended by \citet{1992AJ....103.1159M} and
\citet{2001A&A...370...34R}, but is perhaps preferable given the large
wavelength baseline afforded by the current measurements.  In
addition, it agrees well with the results found above for the
respective reddenings of the blue and red stars determined from
comparison of the IC 10 and SMC CMDs.  For isochrones corresponding to
ages older than 20 Myr we applied a reddening of $E(B-V) = 0.95$ mag.,
as determined from our comparison of the locations of the red stars in
the SMC and IC 10 CMDs.

\begin{figure}
\plotone{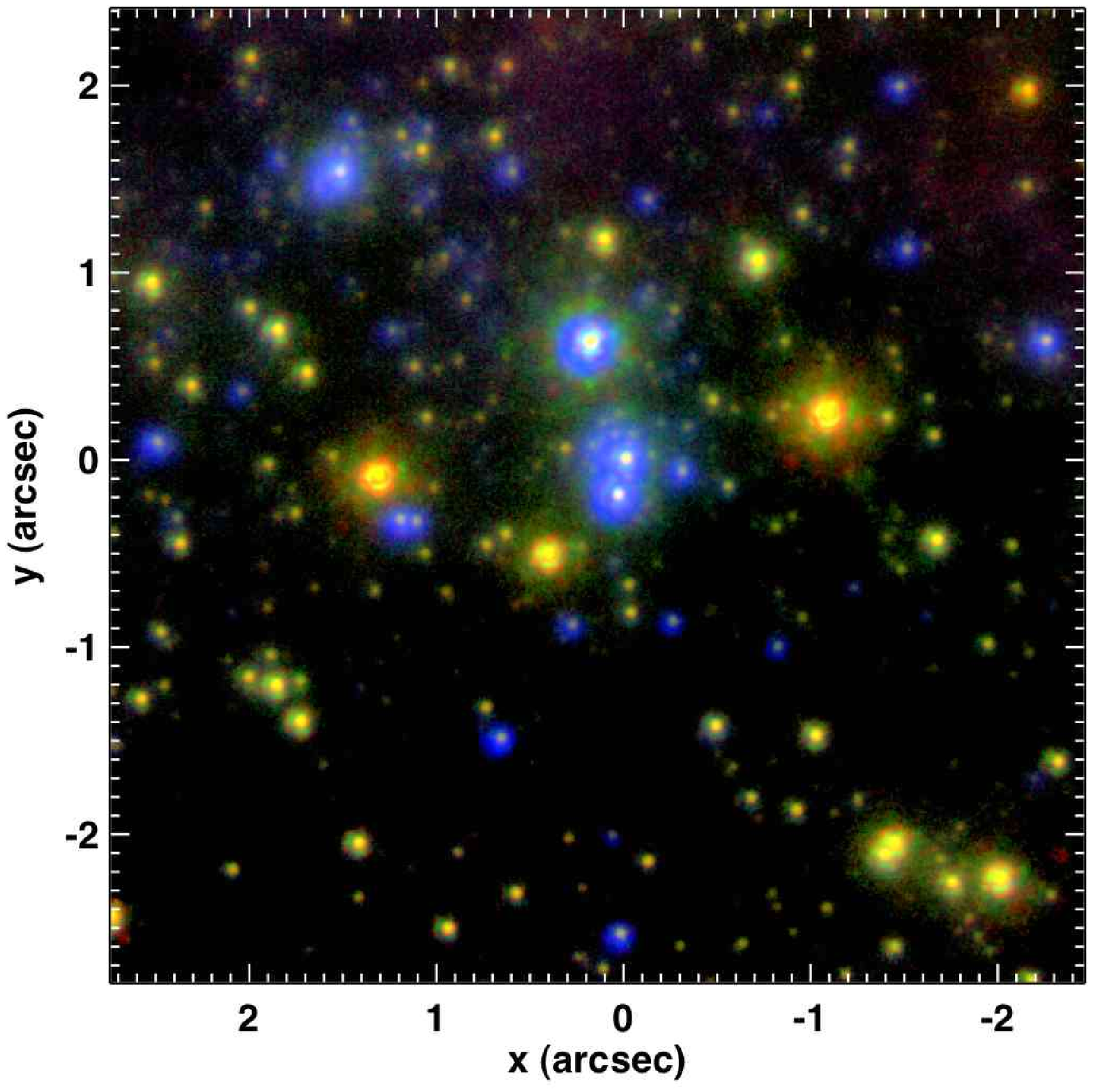}
\caption{ 
An enlargement of the region surrounding [MAC92]~24 from Fig.\
\ref{fig:IHK}, showing that the W-R candidate [MAC92]~24-A
consists of at least three components, two of which (24-A/N and
24-A/S) have nearly equal brightness and blue colors. A third
component 24-A/W ($0\farcs3$ W and $0\farcs1$ S of 24-A/N) is
significantly fainter, but is also blue.  The field is directly
comparable with the WFPC2/F555W image of \citet{2003A&A...404..483C}.
}
\label{fig:IHK_enlarge}
\end{figure}

The comparison between our data and the isochrones indicates that the
[MAC92] 24 region of IC~10 comprises two separate stellar populations:
one group of blue main-sequence stars and a few blue supergiants,
younger than about 10--20 Myr, and a second group of red giants and
AGB stars with ages between 150 and 500 Myr.  This dichotomy suggests
that the recent starburst that produced the W-R stars in [MAC92]~24
sits atop an older field population.  The comparison with the
isochrones draws attention to several stars that are much redder than
predicted by the theoretical models.  As noted above, the bright
objects are C-stars, while the fainter objects are probably highly
evolved AGB stars close to their final luminosity, but extinguished by
optically thick, dusty circumstellar envelopes.

\subsection{Tip of the Red Giant Branch Distance}

The qualitative comparison between the CMDs for the SMC and IC 10
indicates a distance modulus of about 24.5 mag (see \S 3.2).  A
quantitative approach to matching the two CMDs using the
two-dimensional Kolmogorov-Smirnov test \citep{1987MNRAS.225..155F}
proved unsatisfactory because the results were strongly biased by the
incompleteness of the two photometric catalogs. However, we can refine
the distance estimate by using the tip of the red giant branch (TRGB)
method \citep{1993ApJ...417..553L}.  The TRGB marks a strong break in
the luminosity function of the Magellanic Clouds at $K \simeq -6.6$
mag.\ and provides a readily identifiable fiducial point for distance
measurements.

We adopted an empirical approach, comparing the TRGB brightness levels
in the SMC and IC~10 data sets to derive a relative distance modulus
between the two systems. We followed the method developed by
\citet{2000A&A...359..601C} to estimate the location of the TRGB from
the [F814W], $H$, and $K'$ luminosity functions in these two systems.
The luminosity functions and their second derivatives, from which the
location of the TRGB is determined, are shown in Figure \ref{fig:LF}.
Considered individually, the second derivatives of the F814W], $H$,
and $K'$ luminosity functions for IC 10 seem noisy, and the choice of
which peak corresponds to the TRGB ambiguous.  However, simultaneous
consideration of all three wavelengths reveals that there is a common
feature relative to the SMC TRGB (bottom panel of Fig. \ref{fig:LF})
corresponding to a unique relative distance modulus.  Combining the
differential distance moduli from the [F814W], $H$, and $K'$
luminosity functions, using a weighted mean, and correcting for the
fact that the uncertainty in the extinction correction introduces
covariance between these measurements, we find $(m-M)_0 = 24.48 \pm
0.08 $ mag., where we have used the distance and reddening to the SMC
adopted in \S \ref{sec:CMD}.  Because we adopted a large uncertainty
in the reddening towards IC~10 ($E(B-V) = 0.95 \pm 0.15$ mag.), the
$K'$-band TRGB brightness is strongly weighted in this result. Since
the $K'$-band TRGB brightness is sensitive to variations in age and
metallicity, our distance estimate is subject to systematic errors
resulting from any differences in the properties of the red stellar
populations between IC~10 and the SMC.  In order to estimate this
systematic error, we used the results given by
\citet{1998MNRAS.297..872M} for population II giants to determine the
possible systematic uncertainty in the $K'$-band TRGB mag arising from
the width of the [F814W]$-K'$ color of the red giant branch in our CMD
for IC~10. Assuming that the bolometric luminosity at the TRGB is
constant and adopting the $m_{bol} - K$ calibration for population II
giants of \citet{1998MNRAS.297..872M}, we find that any differences in
the the [F814W]$-K'$ color of the red giant branches between the SMC
and IC 10 should produce a systematic error in the true distance
modulus of no more than $\pm 0.16 $ mag. This value includes the
uncertainty in the reddening, $E({\rm [F814W]}-K')$.

We note that our reddening and distance estimates are in good
agreement with those determined by \citet{1996AJ....111..197S} and
\citet{1996AJ....111.1106W} from an analysis of Cepheid data. Our
reddening estimate is about 20\% smaller and our distance is 20\%
larger than the values estimated by \citet{1999ApJ...511..671S}, who
also used Cepheids for their determinations.

\subsection{Wolf-Rayet Stars and Young Clusters}
\label{sec:wr-clu}

An enlargement ($5'' \times 5''$) of the region surrounding [MAC92]~24
is provided in Fig.\ \ref{fig:IHK_enlarge}. This figure can be
compared directly with the WFPC2/F555W image shown by
\citet{2003A&A...404..483C}.  Our observations resolve the W-R source
[MAC92]~24-A into at least three components, two of which (24-A/N and
24-A/S) have nearly equal brightness and blue colors (${\rm [F814W]} -
K' < 0$); a third component, 24-A/W ($0\farcs3$ W and $0\farcs1$ S of
24-A/N), is significantly fainter but is also blue (${\rm [F814W]} -
K' \simeq 0.1$).  The sources [MAC92]~24-A/N and S are the brightest
blue objects in the $K'$ versus [F814W]-$K'$ CMD (Fig. \ref{fig:CMD_compare}).  
Their observed colors and magnitudes
place them above and to the left of the reddened main sequence; their
extinction-corrected near-IR colors ($ ({\rm [F814W]} - K')_0 \simeq
-0.6$ mag.; $(H - K')_0 \leq 0.1$ mag.) and $K$-band luminosities
($M_{K',0} \simeq -6.2$ mag.) are consistent with their being
late-type WN (WN7--9) stars \citep[e.g.,][]{2006mnras...00..000S}.  If
[MAC92]~24-A/N and S are indeed late-type WN stars, they should have
relatively weak lines, which may explain both their absence from the
emission-line survey of \citet{2001A&A...366L...1R} and the weakness
of the He II emission detected by \citet{2002ApJ...580L..35M} in the
spectrum of the blended object observed under seeing-limited
conditions. The W-R candidate [MAC92]~24-B is also resolved into two
blue sources, oriented east-west.  The individual components of
[MAC92]~24-B have colors similar to those of [MAC92]~24-A/N and S, but
are about 2 mag.\ dimmer, which makes them candidate early-type WN
stars. Based on its color and magnitudes, [MAC92] 24-C appears to be a
late O V star. High spatial resolution spectroscopy is needed to
confirm our proposed spectral types for the various components of
[MAC92]~24.

The abundance of W-R stars relative to O stars is a strong function of
time; for an instantaneous burst of star formation and a Salpeter IMF
extending to 100~$M_\odot$, the W-R/O ratio is $\ge 0.1$ only between
3.8--4.7 Myr \citep{1999ApJS..123....3L}.  Our W-R candidates are
drawn from a list of 29 blue stars brighter than the $K'$ mag expected
for a B0V star at the distance and reddening of IC 10.  Therefore, it
would not be surprising if our four robust W-R candidates were
confirmed spectroscopically to be WN stars.  The corollary is that we
can state with confidence that massive star formation in the [HL90]
111c region was active as recently as four Myr ago.

If our proposed W-R candidates are confirmed spectroscopically, the
WC/WN ratio in IC 10 would drop from 1.2--1.3
\citep{2002ApJ...580L..35M,2003A&A...404..483C} to $\sim 1$. While
this value is still anomalously high, given the low metallicity of IC
10, our results suggest that there may be a number of WN stars yet to
be discovered in the rich stellar fields associated with the regions
of recent star formation in this galaxy.

\begin{figure}
\plotone{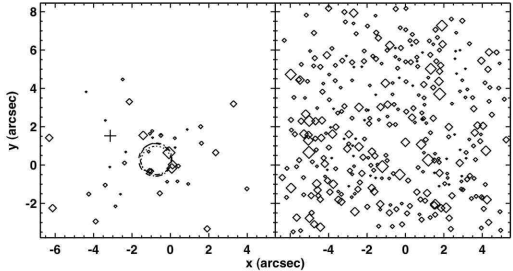}
\caption{Spatial distribution of young ($ < 10$ Myr) blue stars (left)
and older red stars (right).  The young stars are concentrated in the
vicinity of [MAC92]~24.  The three circles (dotted, dashed, dot-dash)
indicate the centroid and half-light radii of the young stars in
[F814W], $H$, and $K'$, respectively.  The half-light radius is about
$0\farcs8$ (3 pc) in all three bands.  The coordinate grid is the same
as in Fig. \ref{fig:IHK} with N up and E to the left, and centered on
the W-R star [MAC92]~24-A/N.  The cross marks the location of cluster
4-1 determined by \citet{2001ApJ...559..225H}.  The largest diamonds
have $K' = 16.5-17.1$ mag., the smallest have $K'=21.9-22.5$ mag.  }
\label{fig:spatdistribs}
\end{figure}

In addition to the stars we have discussed above, there is also a
handful of fainter blue objects clustered within $0\farcs2$ around
[MAC92]~24-A/N. The concentration of luminous blue stars in the region
suggests that the [MAC92]~24 complex could be considered a resolved
stellar cluster, as originally suggested by
\citet{2001ApJ...559..225H} based on inspection of WFPC2 data.
Comparison of the relative spatial distributions of blue and red
stars, as revealed by our F814W and near-IR data and shown in Fig.\
\ref{fig:spatdistribs}, considerably strengthens this proposal.  In
this figure we have divided the stellar population into ``red'' and
``blue'' samples based on their locations relative to the 10 Myr
isochrone in the CMD (Figure \ref{fig:CMD_compare}).  To account for
photometric errors we also include in the blue sample those stars that
lie redward of this isochrone within a swath of width equal to that of
the $1\sigma$ uncertainty in the ${\rm [F814W]} - K'$ color.

With this definition, the red stars appear to be far more uniformly distributed
across the field than the blue stars, which are concentrated around a
position centered near [MAC92] 24 A and clearly delineate a stellar
cluster. For stars with ages $\leq 10$ Myr, the luminosity weighted
centroid is located $0\farcs8$ E of [MAC92] 24-A/N, and the half-light
radius is $0\farcs83$ (3.2 pc) in all three bands. The integrated
magnitudes within the half-light radii of the cluster are [F814W] =
17.07 mag., $H$ = 16.71 mag., and $K'$ = 16.48 mag.  The half-light
radius grows steadily with the age cut-off. For an age of $< 155 $ Myr
we find half-light radii of $1\farcs22$ (4.7 pc) in F814W and an
integrated magnitude of [F814W] = 16.47 mag., which is in good
agreement with the measurements of \citet{2001ApJ...559..225H}.

Based on PSF fitting, only one source in our field has a spatial
profile that is significantly more extended than that of a point
source. This object is identified as number 7 in Figures
\ref{fig:kprimefindingchart} and \ref{fig:CMD}.  From its deconvolved
FWHM we infer a size of only $0\farcs05$ (0.2 pc).  It is therefore
probably too small to qualify as a cluster and is more likely an
unresolved association of a few massive stars.

\subsection{Nebular Emission}

The three-color image (Fig.\ \ref{fig:IHK}) reveals a swath of diffuse
emission, whose extent and location correlate extremely well with that
seen in the archival WFPC2/F656N image of this field.  As F656N traces
H$\alpha$ $\lambda$ 6563 we believe that this extended component
represents recombination emission from ionized gas.  This suggestion
is supported by the morphology of the emission, which appears to form
an arc around [MAC92]~24.  The emission in the $H$ and $K'$ bands is
likely a combination of free-free, bound-free and Brackett lines.
Although the F814W filter excludes H$\alpha$ $\lambda$ 6563 and [NII]
$\lambda\lambda$ 6547, 6583, it does include [S II] $\lambda\lambda$
6716, 6731, [S III] $\lambda\lambda$ 9069, 9532 and the series Paschen
limit.  We estimated the flux of this diffuse component from our
star-subtracted images; we find fluxes of $0.7$, $1.5$, and $2.0$ mJy
in the F814W, $H$, and $K'$ bands, respectively.

In ionization equilibrium the observed nebular flux is approximately
proportional to the Lyman continuum flux and therefore can be used to
place a constraint on the number of hot stars. In order to determine
the conversion between the nebular emission and the Lyman continuum
flux, we used the MAPPINGS 3 code \citep{1993ApJS...88..253S} to
construct the nebular emission spectrum from an ionization-bounded HII
region excited by a source with a color temperature of 30,000 K. We
find that a source that produces $10^{51}$ ionizing photons per second
at a distance $(m-M)_0 = 24.5$ yields equivalent flux densities of
6.8, 3.3 and 4.3 mJy in the F814W, $H$ and $K'$ filters, respectively.

Our measurements imply a total ionizing flux of $3-6 \times
10^{50}$~s$^{-1}$, where we have adopted our previously measured
extinction to the blue stars of $E(B-V) = 0.6 $ mag. With the
effective temperature and ionizing flux calibrations of
\citet{2006ApJ...638..409H} for O stars in the SMC cluster NGC 346, we
find that this ionizing flux corresponds to $\sim$ 50--100 SMC O6.5V
stars.  With an age of about 3 Myr and a metallicity of about one
fifth solar, the young SMC cluster NGC 346 is an appropriate and
useful analog for the [MAC92] 24 region.  Conversion from $N_{O6.5V}$
to a total number of O stars requires assumptions about the age and
initial mass function of the stellar population. We again used the
results of \citet{2006ApJ...638..409H} for the O stars in NGC 346 and
fit the ionizing flux as a function of evolutionary mass in order to
estimate the ratio $N_{\rm O6.5V}/N_{\rm O V}$
\citep[see][]{1994ApJ...421..140V}.  For a Salpeter IMF with O star
masses between 20 and 85 $M_\odot$, we find that this ratio has a
value of approximately 2. This indicates that 30--50 SMC-like main
sequence O-type stars are needed to power the nebular emission seen in
our image of the [MAC92] 24 region of IC 10.  The ratio $N_{\rm
O6.5V}/N_{\rm O V}$ depends on the mass range used; if the upper mass
limit is reduced to 40 $M_\odot$, then the ratio drops to about unity.
Fig.\ \ref{fig:CMD} reveals $\sim 19$ blue stars bright enough to be
O8 stars or earlier ($K < 20.8$ mag.). Four of these stars are WN
candidates.  The correspondence between the expected and observed
number of hot stars is satisfactory, given the potential for
systematic errors in estimating the nebular emission from broad-band
filters and the uncertainties in our analysis.  This agreement
suggests that we have identified the dominant source of the ionizing
flux in the region.

\section{Conclusions}

We have combined ground-based laser guide star adaptive optics and
{\it Hubble Space Telescope} imaging of the local group dwarf
starburst galaxy IC~10 to yield a high resolution view of the stellar
population.  These data resolve the Wolf-Rayet candidate [MAC92]~24
into multiple components.  The object [MAC92]~24A is found to be
composed of at least three bright, blue sources while the object
[MAC92]~24~B is similarly resolved into two blue sources. The
luminosities and colors of these component sources are consistent with
their being massive hot stars and we propose that there are at least
four robust WN candidates in the field.  Integral field AO
spectroscopy will be necessary to confirm the status of these objects.

The $K'$ versus ${\rm [F814W]}-K'$ CMD generated from our data
exhibits a morphology similar to that found for the SMC, which has a
similar metallicity.  Comparison of the IC~10 CMD with theoretical
isochrones indicates that star formation in the vicinity of [MAC92]~24
has generated two distinct stellar populations, one relatively young
($\sim 10$~Myr) and one substantially older (150--500~Myr).  A clear
spatial segregation between the young blue stars (W-R stars and
main-sequence OB stars) and the older, red stellar population is
apparent in our images.  The blue stars in our IC 10 field are
clustered into a distinct OB association with a half light radius of
about 3 pc. The ionizing flux from this resolved young population is
responsible for a large fraction of the nebular continuum observed in
this region.  The red stars, comprising red giants, O- and C-rich
asymptotic giant branch stars and highly evolved stars with thick
circumstellar envelopes, are much more uniformly distributed.  Spatial
segregation also occurs along the light of sight, with a differential
reddening of $E(B-V)$ of about 0.4 mag.\ between the red and the blue
populations.  The column of neutral gas in IC ~10 towards this
direction is $N_{HI + H_2} = 2.2 \times 10^{21}$ cm$^{-2}$
\citep{1998AJ....116.2363W, 2006ApJ...643..825L}.  Thus, the blue
stars, with $E(B-V) \sim 0.6$ mag., appear to be on the front surface
of IC~10, while the red stars, which have $E(B-V) = 0.95$ mag., probe a
significant fraction of the column depth of the IC 10 interstellar
medium.

Finally, comparison of the luminosity functions for red stars in IC 10
and the SMC yields a tip-of-the giant branch distance to IC 10 of
$(m-M)_0 = 24.48\pm 0.08$ mag.\ with a systematic uncertainty of $\pm
0.16$ mag.

\begin{acknowledgements}

We thank J.\ Anderson and M.\ Sirianni for discussions regarding ACS
photometry.  We would like to thank M.\ D.\ Perrin for maintaining the
NIRC2 data reduction pipeline.  This work has been supported by the
National Science Foundation Science through AST 0205999 and the Center
for Adaptive Optics, managed by the University of California at Santa
Cruz under cooperative agreement No. AST 9876783.  The W.M. Keck
Observatory is operated as a scientific partnership among the
California Institute of Technology, the University of California, and
the National Aeronautics and Space Administration.  The Observatory
was made possible by the generous financial support of the W.M. Keck
Foundation.  Finally, the authors wish to recognize and acknowledge
the very significant cultural role and reverence that the summit of
Mauna Kea has always had within the indigenous Hawaiian community.  We
are most fortunate to have the opportunity to conduct observations
from this sacred mountain.

\end{acknowledgements}

\begin{table}
\caption{Photometry$^a$}
\begin{center}
\begin{tabular}{rcrrrrr} 
\hline
\hline
ID & Name & $x^b$ & $y^b$ & [F814W] &  $H$ &$K'$ \\
   &      & (arcsec) & (arcsec) & (mag.) & (mag.) & (mag.)  \\ 
\hline 
%
%
%
1  & $\ldots$  & +0.612  & +1.590  & 21.76  & 22.61  & 22.29 \\
2  & (24-B/W)  & +1.121  & --0.337  & 20.33  & 20.89  & 20.73 \\
3  & (24-A/S)  & +0.041  & --0.194  & 17.85  & 18.38  & 18.15 \\
4  & (24-A/N)  & 0.000  & 0.000  & 18.26  & 18.51  & 18.34 \\
5  & (24-B/E)  & +1.204  & --0.326  & 20.33  & 20.35  & 20.28 \\
6  & (24-A/W)  & --0.301  & --0.070  & 20.92  & 20.80  & 20.81 \\
7  & $\ldots$  & +1.531  & +1.528  & 19.02  & 19.03  & 18.78 \\
8  & (24-C)  & +0.666  & --1.493  & 21.14  & 20.75  & 20.67 \\
9  & $\ldots$  & +0.187  & +0.624  & 17.70  & 16.99  & 16.74 \\
10  & $\ldots$  & --1.671  & +3.687  & 18.35  & 16.73  & 16.50 \\
11  & $\ldots$  & +6.086  & +4.707  & 20.67  & 17.97  & 17.64 \\
12  & $\ldots$  & --1.788  & +7.255  & 20.91  & 18.01  & 17.45 \\
13  & $\ldots$  & --2.006  & --2.247  & 21.32  & 18.26  & 17.78 \\
14  & $\ldots$  & +5.171  & +1.269  & 21.77  & 18.45  & 17.86 \\
15  & $\ldots$  & +5.125  & +2.281  & 21.39  & 18.04  & 17.44 \\
16  & $\ldots$  & --1.207  & +5.004  & 21.10  & 17.62  & 17.12 \\
17  & $\ldots$  & +0.409  & --0.514  & 21.88  & 18.30  & 17.69 \\
18  & $\ldots$  & --1.082  & +0.241  & 21.17  & 17.51  & 16.71 \\
19  & $\ldots$  & +1.318  & --0.099  & 21.81  & 17.92  & 17.01 \\
20  & $\ldots$  & +5.136  & +0.643  & 21.77  & 17.20  & 16.53 \\
21  & $\ldots$  & --2.150  & +1.966  & 24.54  & 20.02  & 19.25 \\
22  & $\ldots$  & +3.142  & +0.829  & 24.79  & 20.23  & 19.34 \\
23  & $\ldots$  & --4.077  & +0.728  & 24.27  & 19.13  & 18.25 \\
\hline
\end{tabular}
\end{center}
\label{TAB:phottable}
a) For errors see \S\S \ref{SEC:ir-data} \& \ref{SEC:hst-data} \\
b) Relative to [MAC92] 24-A/N at 
RA(2000.0) $00^h$ $20^m$ $27^s.36$, DEC(2000.0) 
$+59^\circ$ $17'$ $37\farcs33$.
\end{table}

\clearpage

\end{document}